\documentclass[preprint]{elsarticle}
\usepackage{amsmath, bm}
\usepackage{hyperref}
\usepackage{subfig}
\usepackage{lineno}

\long\def\ignore#1{\relax}

\makeatletter
\def\ps@pprintTitle{%
  \let\@oddhead\@empty
  \let\@evenhead\@empty
  \def\@oddfoot{\reset@font\hfil\thepage\hfil}
  \let\@evenfoot\@oddfoot
}
\makeatother

\title{Response to Comment from Robert Cousins on Confidence intervals for the Poisson distribution\tnoteref{t1}}
\tnotetext[t1]{Work supported in part by the U.S. Department of Energy, Office of Science, Office of High Energy Physics, under Award Number DE-SC0011925.}
\author{Frank C.~Porter\corref{cor1}}
\address{Physics Department 356-48, California Institute of Technology, Pasadena, CA 91125 USA}
\cortext[cor1]{Corresponding author, fcp@caltech.edu}
\ead{fcp@caltech.edu}
\date{\today}

\begin{document}

\begin{abstract}
Robert Cousins has posted a comment on my manuscript on ``Confidence intervals for the Poisson distribution''. His key point is that one should not include in the likelihood non-physical parameter values, even for frequency statistics. This is my response, in which I contend  that it can be useful to do so when discussing such descriptive statistics. 
\end{abstract}


\maketitle



Cousins~\cite{cousins2025commentfrankporterconfidence} criticizes my study of Poisson confidence intervals~\cite{porter2025confidenceintervalspoissondistribution}, objecting to my use of the likelihood function in regions of parameter space which are deemed non-physical. His point is that the likelihood function should be considered undefined in such a region, because the physical model is not defined there. In this note I respond to this criticism. I suspect that we will not be able to agree, but it seems useful to expand on the discussion in my paper by responding directly on this point.

While the issue is general, we may have in mind the concrete situation of~\cite{porter2025confidenceintervalspoissondistribution}, where we sample random variable (RV) $N=0,1,2,\ldots$ from a Poisson distribution with mean $\mu$ equal to the sum of a ``signal'' strength $\theta\ge 0$ and a known background $b$ (also $\ge 0$):
\begin{equation}
f(n;\theta,b) = \frac{\mu^n}{n!}e^{-\mu}=\frac{(\theta+b)^n}{n!}e^{-\theta-b},
\label{eq:Poisson}
\end{equation}
 where $n$ is a possible value of $N$.

It is very important to keep in mind that the entire discussion is in the context of frequency statistics.
In any alternative discussion involving degree-of-belief, in particular Bayesian statistics, there is no disagreement.
Even if I try to define a likelihood function in a non-physical region of parameter space, the physical region is enforced by the prior. The belief that the parameter is non-physical is zero, complete agreement.

However, when we are trying to describe the measurement, and not the parameter, the situation is quite different. When we provide the result of a Poisson sampling, we give $n$ as the sampled value of
random variable $N$. We know that $n$ is the maximum likelihood estimator for the Poisson mean $\mu$. For a parameter $\theta = \mu-b$, it is then straightforward to describe the measurement relevant to $\theta$ as $\hat\theta=n-b$. Of course, $n$ might happen to be less than $b$ in some samplings. As a description, it is less informative to say the measurement in that case is 0. To put it more technically, $N-b$ is an sufficient statistic for $\theta$.  On the other hand, $\max(N-b, 0)$ is not a sufficient statistic. 
																	
It is stated in~\cite{porter2025confidenceintervalspoissondistribution} that, while our descriptive statistics are not inferences, they can provide useful information towards forming an inference.
In connection with my discussion that frequency statistics are descriptive only and cannot be interpreted as degree-of-belief statements, Cousins suggests that there is also an inferential purpose to frequency statistics. He doesn't expand on this, but I
would agree in the following sense:
If a measurement yields the result $\hat\theta = n-b$, then we reasonably infer that 
$\theta$ is ``more likely'' to be near  $\hat\theta$ than far from it. In doing so, we have entered the
domain of degree of belief. If we know on physical grounds that $\theta\ge0$, and we observe a negative value of  $\hat\theta$, then we think that $\theta\ge 0$ is probably (degree of belief) pretty small (but of course not negative), in order for
such a negative fluctuation to be likely.

Cousins~\cite{cousins2025commentfrankporterconfidence} argues that it is improper to define a likelihood function that is non-zero in a physically forbidden region of parameter space. Statistics books tend to be rather silent on the issue. Shao~\cite{Shao2003} has an example that supports Cousins' point of view. However, it appears in a chapter with heavy Bayesian content, and Shao does not elaborate on the context.
In any event, it is such a useful tool that, should some statistical semantic authority ban such usage, I
would just call it something else (``descriptionhood'' and ``maximum descriptionhood'', maybe). We can use the concept without violating any physical model (e.g., the knowledge that $\theta\ge 0$). 
The discussion above about the description by $\hat\theta = n-b$ is, of course, just what we get when we maximize the likelihood
\begin{equation}
L(\theta;n) = \frac{\mu^n}{n!}e^{-\mu}=\frac{(\theta+b)^n}{n!}e^{-\theta-b},
\end{equation}
which is mathematically well-defined for any $\mu\ge0$.
We are making no suggestion that the probability model is anything other than $\theta\ge 0$. We are only allowing 
$\theta$ to enter into the negative region in the likelihood function in order to describe a measurement. The sampling distribution is always assumed to be with $\theta\ge 0$. 

This discussion has been in the context of the Poisson distribution, but the same considerations arise for other sampling distributions. For example the situation for the
normal distribution is discussed in~\cite{NarskyPorterNormal}. 

Cousins~\cite{cousins2025commentfrankporterconfidence} also takes issue with my stance that frequentist statistics should be regarded as descriptive, potentially useful for inferential statements, but should not be regarded themselves as providing such statements. I acknowledge that I may be taking a more extreme position than considerable historical precedent. However,  this position has served well to keep distinctions straight. Thus, for example, I am perfectly fine with describing a measurement with a downward fluctuation
as $\hat \theta = -2\pm 1$ when I know that in truth $\theta>0$. The inference that I  draw from the description is that ``probably'' $\theta$ is not much larger than zero. Unfortunately, the meaning of ``probability'' in such an inferential statement is rather vague.
It is up to Bayesian methodology to provide an interpretation in terms of degree of belief.

Cousins cites the classic statistics work by Kendall et al.~\cite{Kendall1999} in this discussion of inference. Volume 2A discusses several approaches to estimation, including Bayesian. The authors recognize the issues (and controversies) associated with these approaches. In particular, they summarize their views in a discussion in sections 26.58 through 26.78. While this discussion is somewhat nuanced, I find my point of view consistent with it. Section 26.69 states, for example, ``\dots the failure (?) of the frequency approach to deliver statements on the credibility of a hypothesis is almost axiomatic, since frequentists are unwilling to accept any probability$_1$ concepts that do not have a frequency interpretation.'' Note that ``probability$_1$'' is here defined as degree of belief. 

The insistence in pointing out that frequency statistics are strictly descriptive is motivated by the abundance of confusion resulting when attempts are made to interpret them as intended for something else. People have even claimed that frequency statistics is ``wrong'' because of this confusion. A prominent example from a decade ago occurred when the editors of a psychological journal claimed that $p$ values (which are manifestly frequency statistics) are ``invalid'' and decided to ban them in their journal~\cite{Trafimow02012015}.

I have explained my take on the main points raised by Cousins in~\cite{cousins2025commentfrankporterconfidence} . While I don't anticipate universal agreement, I claim that my perspective is both coherent and useful.

\bigskip

\section*{Acknowledgements}

Work supported in part by the U.S. Department of Energy, Office of Science, Office of High Energy Physics, under Award Number DE-SC0011925. I am grateful to Robert Cousins for a helpful comment.

\bigskip

\bibliographystyle{unsrt}
\bibliography{RCresponse}

\end{document}